\documentclass[final,3p,times]{elsarticle}
\usepackage{amssymb}
\usepackage{amsmath}
\input{tcilatex}
\journal{Zeitschrift 	f$\ddot{u}$r Angewandte Mathematik und Mechanik}

\begin{document}

\begin{frontmatter}

 \title{Symmetric forms for   hyperbolic-parabolic systems
 	of multi-gradient fluids}

\author[Henri]{Henri Gouin }
\ead{henri.gouin@univ-amu.fr;henri.gouin@yahoo.fr}
\cortext[cor1]{Corresponding author.}\corref{cor1}

\address[Henri]{Aix-Marseille Univ,  CNRS, IUSTI 
 UMR 7343,
13013 Marseille, France}

\begin{abstract}
	$\ddot{a}$
 We consider multi-gradient fluids endowed with a volumetric internal energy which is a function of   mass density, volumetric entropy
 and their successive  gradients. We
 obtained  the thermodynamic forms of    equation of motions and   equation of energy, and
the motions  are compatible with the two laws of
 thermodynamics.\newline
 The
 equations of multi-gradient fluids belong to the class of dispersive systems. In the  conservative case,  we can replace the  set
 of   equations by a   quasi-linear system  written in a
 divergence form.    Near an equilibrium position, we obtain a 
 symmetric-Hermitian system of equations in the form of Godunov's systems.  The equilibrium positions are proved to be stable when the total volume energy of the fluids is a convex function with respect  to convenient conjugated variables - called main field - of mass density, volumetric entropy, their successive gradients, and velocity.
\end{abstract}


\end{frontmatter}

\section{Introduction}
In continuum mechanics, Cauchy has only described the behavior of a mechanical system when inhomogeneities have a characteristic length scale much smaller than the macro-scale in which the phenomena are observed. Usually, the mechanical description of other conservative systems needs a higher order stress tensor and they are many physical phenomena described by this generalized continuum theory. For instance Piola’s continua need a \textit{(n+1) -- uple} of hyper stress-tensors where the order is increasing from  second to  \textit{n+1}; the contact interactions do not reduce to forces per unit area on boundaries, but include $k$-forces  concentrated on areas, on lines or even in wedges \cite{Isola1,Isola2}. 
The \textit{ (n+1) -- th} order models are suitable for describing non-local effects as in   bio-mechanical phenomena \cite{Madeo,GouinHR}, damage phenomena   \cite{Yang}, and internal friction in solids \cite{Limam}. 
The range of validity of Noll’s theorem  is not verified but the second principle of thermodynamics is clearly proved \cite{Toupin,Mindlin}. Many   efforts have been made to study these media, theoretically and numerically, where the research of symmetric forms for the equations of processes must be a main subject to verify the well-posed  mathematical problems.

 For fluids,  across    liquid-vapor interfaces,   pressure  $p$  and    volumetric internal energy $\varepsilon_{0}$
are
non-convex functions of   volumetric entropy $\eta$ and  mass density.
Consequently, the simplest continuous model allowing to study non-homogeneous
fluids inside interface layers considers another volumetric internal energy
$\varepsilon$ as the sum of two terms: the first one  
defined as $\varepsilon_{0}$ and the second one 
associated with the non-uniformity of the fluid which is approximated by an
expansion limited at the first gradient of mass density.
This form of energy  which  describes interfaces
as diffuse layers  was first  introduced by  van der Waals
\cite{Waals} and  is now widely used in the literature \cite{Cahn}. The model has many applications  for inhomogeneous fluids   \cite{Gouin1, Gouin2} 
and  is extended  for  different materials
in
continuum mechanics,
which modelizes the behavior of
strongly inhomogeneous media
\cite{garajeu,GouinHR,Gavrilyuk,Gouin-Ruggeri,Eremeyev,HMR}. 
The model  
yields a  constant temperature   at equilibrium.
Consequently,
the volume entropy  varies with the
mass density   in the same way as in the bulks.   This  first assumption of van der
Waals   using long-ranged but weak-attractive
forces   is not exact for realistic intermolecular potentials
and the thermodynamics  
is not completely considered \cite{Evans}. 

For variational principles, it
is not possible to take directly the temperature gradient  into account  : the volume internal energy must be a functional   of  canonical variables, \textit{i.e.} mass density and volumetric entropy.    
The simplest
model was called \emph{thermocapillary fluid model}  when the internal energy depends on mass density, volumetric entropy and their first gradients \cite{casal4, Rowlinson}.  Such a 
behavior has also been considered in   models when at
equilibrium the temperature is not constant  in 
inhomogeneous parts of complex media
\cite{Maitournam,Forest,Seppecher}.
\newline
To improve the model accuracy, the  general case considers fluids when the volume
internal energy depends on  mass density, volumetric  entropy and their
gradients up to a  convenient $n$-order ($n \in \mathbb{N}$) where
continuum models  of
\emph{gradient theories}   are useful in  case of strongly inhomogeneous fluids \cite{Germain1,Seppecher2}. The  models  have a justification in the framework of  mean-field molecular theories  when the van der Waals forces
exert stresses on fluid molecules producing surface tension effects  \cite{Evans,Rowlinson,Widom,Saccomandi}.
\\
In \cite{Gouin}, we obtained  the equation of
motions for perfect multi-gradient fluids.
For dissipative motions, the
conservation of mass and   balance of entropy implied the
equation of energy.   The Clausius-Duhem inequality
was deduced from   viscous-stress dissipation and  Fourier's
equation.\\

Moreover, the symmetrization of the equations of mechanical systems is a main subject of study for the structure of solutions of complex media, and being
still debated.  \\

First, we   present some historical remarks which are detailed in Ref. \cite{RS} :\\
In 1961, Godunov wrote a  paper on \textit{an interesting class of quasi-linear sytem} which   proves   that with convenient change of variables, the
 system of Euler fluids becomes symmetric. He also proved that all systems coming from  variational principles can be written in   symmetric form \cite{Godunov}.
In 1971, Friedrichs and Lax proved that all systems  compatible with the entropy principle are symmetrizable \cite{Friedrichs}:    after a pre-multiplication with a convenient matrix,   systems   become   symmetric.  
In  1974, Boillat introduced a  new field   of variables  for which the original system can be written in a symmetric form \cite{Boillat}. He was the first  who symmetrized  original hyperbolic systems   that were compatible with the entropy principle. He  called the systems, \textit{Godunov's systems}   \cite{LectureNotes}.
The technique of Lagrange multipliers to study the entropy principle was given first by I-Shi Liu \cite{Liu}, and was similar to the work by Ruggeri and Strumia which were interested in extending the previous technique to  the relativistic case by using a covariant formulation \cite{Strumia}.
In  1982, Boillat extended the symmetrization  to the case with constraints \cite{Boillat2}; the problem was also considered by Dafermos \cite{Dafermos}.
In 1983,  
Ruggeri realized it was possible to construct a symmetrization  for parabolic sytems and he wrote down  the expression of \textit{the main field  of variables} for Navier-Stokes-Fourier fluids \cite{Acta_Ruggeri}, and  in 1989, he proved that symmetrization was compatible with the Galilean invariance \cite{Ruggeri3,Ruggeri4}.

Second, we consider the framework of    models which are represented by quasi-linear first-order systems of $%
n$ balance laws (we adopt the sum convection on  repeated indexes)  :
\begin{equation}
\frac{\partial\boldsymbol{G}^{\it 0}(\boldsymbol{v})}{\partial t}+\frac{\partial%
	\boldsymbol{G} ^{j}({\boldsymbol{v}})}{\partial x^{j}}=\boldsymbol{g(\boldsymbol{v})},
\label{sh}
\end{equation}
with an additional scalar balance equation  corresponding to the energy equation in
pure mechanics or the entropy equation in thermodynamics :
\begin{equation*}
\frac{\partial{h}^{\it 0}(\boldsymbol{v})}{\partial t}+\frac{\partial h ^{j}(%
	\boldsymbol{v})}{\partial x^{j}}=  {\it{\Sigma}} (\boldsymbol{v}),  \label{shsu}
\end{equation*}
where $\boldsymbol{G}^{\it 0},\boldsymbol{G}^j, \  j \in \{ \textit{1}, \dots , n\}  $,  $%
\boldsymbol{g}, \boldsymbol{v}$ are column vectors of $\mathbb{R}^n$, and  ${h}^{\it 0}, h ^{j},\,   j \in \{\textit{1},  \dots, n \},\, {\it\Sigma}$\, are scalar functions; scalar $t$, and $ 
\boldsymbol{x}  =  ( x^{\it { 1}}, \cdots , x^{n})$ are  time and $\mathbb{R}^n$-space
coordinates, respectively. 
Function $h^{\it 0}$ is assumed to be convex with respect to field $%
\boldsymbol{G}^{\it 0} (\boldsymbol{v})\equiv \boldsymbol{v}$  %
(see Refs. \cite{Godunov,Friedrichs,Ruggeri2}).  
Dual-vector field $\boldsymbol{v}^\prime$, associated with Legendre
transform $h^{\prime {\it 0}}$ and potentials $h^{\prime j}$ is such that (see Ref. \cite{Boillat}) :
\begin{equation*} 
\boldsymbol{v}^{\prime}= \left(\frac{\partial h^{\it 0}}{\partial \boldsymbol{v}}\right)^\star, \qquad
h^{\prime {\it 0}} = \boldsymbol{v}^{\prime\star} \, \boldsymbol{v}- h^{\it 0}, \qquad
h^{\prime j} = \boldsymbol{v}^{\prime\star}\, \boldsymbol{G}^j(\boldsymbol{v})- h^j, \label{MF} 
\end{equation*}
where $^\star$ indicates the transposition. By a convexity argument, it is possible to take $\boldsymbol{v}^{\prime}$ as a vector field and we obtain :
\begin{equation}  \label{change}
\boldsymbol{v} =\left( \frac{\partial h^{\prime {\it 0}}}{\partial \boldsymbol{v}^\prime}\right)^\star
, \qquad \boldsymbol{G}^j(\boldsymbol{v})= \left(\frac{\partial h^{\prime j}}{%
	\partial \boldsymbol{v}^\prime}\right)^\star.
\end{equation}
Inserting new variables given by Eqs. \eqref{change} into System \eqref{sh}, we get :
\begin{equation*}  \label{symform}
\frac{\partial}{\partial t}\left(\frac{\partial h^{\prime {\it 0}}}{\partial
	\boldsymbol{v}^\prime}\right) + \frac{\partial}{\partial x^j}\left(\frac{%
	\partial h^{\prime j}}{\partial \boldsymbol{v}^\prime}\right) = \boldsymbol{g
}(\boldsymbol{v}^{\prime}),
\end{equation*}
which is symmetric and equivalent to
\begin{equation}
\boldsymbol{A}^{\it 0}\,\frac{\partial \boldsymbol{v}^\prime}{\partial t}+%
\boldsymbol{A}^{j}\frac{ \partial \boldsymbol{v}^\prime}{\partial x^{j}}=%
\boldsymbol{g}(\boldsymbol{v}^\prime),  \label{symm}
\end{equation}%
where matrix $\boldsymbol{A}^{\it 0}\equiv \left(\boldsymbol{A}^{{\it 0}}\right)^\star  $ is 
positive-definite symmetric and matrices $\boldsymbol{A}^{j}=\left(\boldsymbol{A}%
^{j}\right)^{\star }$ are symmetric,
\begin{equation}  \label{matrici}
\boldsymbol{A}^{{\it 0}}\equiv \left(\boldsymbol{A}^{\it 0}\right)^\star= \frac{\partial^2
	h^{\prime {\it 0}}}{\partial \boldsymbol{v}^{\prime 2}}, \qquad \boldsymbol{A}^j\equiv \left(\boldsymbol{A}%
^{j}\right)^{\star }= \frac{%
	\partial^2 h^{\prime j}}{\partial \boldsymbol{v}^{\prime 2}}, \quad (j=\textit{1}, \dots,n). 
\end{equation}
The symmetric form of governing equations implies hyperbolicity. For
conservation laws with vanishing production terms, the hyperbolicity is
equivalent to the stability of constant solutions with respect to
perturbations in form $\ e^{i(\boldsymbol{k}^{\star }\boldsymbol{x}-\omega
	t)} $,\ where $i^{2}=-1,\ \boldsymbol{k}^{\star }=[k_{\it 1},\cdots ,k_{n}] \in   \mathbb{R}^{n  \star} $ and
$\omega$ is a real scalar. Indeed, the symmetric form of governing equations
for an unknown vector $\boldsymbol{v}, \ (\boldsymbol{v}^\star = [v_1,\cdots
,v_{n}]$) implies the \emph{dispersion relation :}
\begin{equation*}
{\rm{det}}\,(\boldsymbol{A}_{(k)}-\omega \boldsymbol{A}^{\it 0})=0 \quad \mathrm{%
	with} \quad \boldsymbol{A}_{(k)}=\boldsymbol{A}^{j}k_{j}\,,  \label{eigenval}
\end{equation*}%
which determines real values of $\omega $ for any \emph{real wave vector\,}$%
\boldsymbol{k}$, where operator\ det\ denotes the determinant. In this case, phase velocities are real and coincide with
the characteristic velocities of hyperbolic system \cite{RMSeccia,BLR}.
Moreover, right-eigenvectors of $\boldsymbol{A}_{(k)}$ with respect to $%
\boldsymbol{A}^{\it 0}$ are linearly independent and any symmetric system is also
automatically hyperbolic.
Symmetric form given by Eq. \eqref{symm} with relations  \eqref{matrici} are commonly called \emph{Godunov's systems} \cite{Godunov}.
\\
In the case of systems with parabolic structure (\emph{hyperbolic-parabolic
	systems}), a generalization of symmetric system   is written : 
\begin{equation}
\boldsymbol{A}^{\it 0}\,\frac{\partial \boldsymbol{v}^\prime}{\partial t}+\boldsymbol{A}%
^{j}\frac{\partial \boldsymbol{v}^\prime}{\partial x^{j}}
-\frac{\partial}{\partial
	x^j} \left(\boldsymbol{B}^{jl}\frac{\partial \boldsymbol{v}^\prime}{\partial x^l }%
\right)=0,  \label{symmPara}
\end{equation}
where matrices\ $\boldsymbol{B}^{jl}= \left(\boldsymbol{B}^{jl}\right)^\star$\   are symmetric, and $
\boldsymbol{B}_{(k)}=\boldsymbol{B}^{jl} k_j k_l $ \ are
non-negative definite.
\newline
The compatibility of hyperbolic-parabolic
systems given by Eq. \eqref{symmPara} with entropy principle and the
corresponding determination of main field is given  in \cite%
{Acta_Ruggeri} for Navier-Stokes-Fourier fluids and in general case  
in \cite{Kawa}. The same authors 
considered linearized version of  
System \eqref{symmPara}  proving that the
constant solutions are stable.   \\

These reminders being given, the aim of  present paper is to extend the results of symmetrization for\textit{ the most general case of multi-gradient  fluids}. Using a
convenient change of variables -- \textit{the main field }--  associated with a Legendre's
transformation of
the total fluid energy,  equations of processes  can be written in this special divergence form  as in Eq. \eqref{symmPara}.   Near an equilibrium position, we obtain a new 
	Hermitian-symmetric form  of the system of perturbations. The
obtained set   belongs to the class of dispersive systems.\\
 The paper is organized as follows :    In Section 2,  we recall the main results obtained in \cite{Gouin} (equations of
conservative motions, balance of energy and compatibility with the two laws of thermodynamics). We additively obtain the existence of a stress tensor which can    write the equation of
motions in a form similar  to those of continuous media. 
 In Section 3,  \textit{the main field of variables} -- for which the conservative equations of motions are written in divergence form -- is obtained. 
In Section 4,    the  Hermitian-symmetric form   for the equations of perturbations near an equilibrium position is deduced.  The perturbations are stable in domains where   the total volume   energy is a convex function of the main field of variables, which     proof confirms that the mathematical problem is well posed.    A conclusion    ends
the paper.

  \section{Multi-gradient fluids and equation of motions}
  In this  section we  recall in a new presentation, \textit{adapted for symmetric calculations}, the main results obtained in \cite{Gouin}, but  subsection {\it 2.3} introduces new calculations allowing to obtain the stress tensor of   conservative multi-gradient fluids.   In this Section,  for the sake of simplicity, we identify vectors and covectors and we always indicate indexes in subscript
  position without taking account of  the tensors' covariance or
  contravariance.  
  \subsection{Definition of multi-gradient fluids}
  We consider perfect fluids
  with a volume internal energy $\varepsilon $ function
  of   volumetric entropy $\eta $,  mass density  $\rho $, and their  
  gradients until order $  n \in \mathbb{N}$, 
  \begin{equation*}
  \varepsilon =\varepsilon (\eta , \rho, \nabla \eta , \nabla \rho , {\ldots },\nabla
  ^{n}\eta ,\nabla ^{n}\rho ) ,
  \end{equation*}%
  where operators $\nabla ^{p}$,\  $p\in \{1,\ldots ,n\}$,\ denote  the
  successive gradient  in Euclidian space ${\mathcal D}_{t}$, of Euler variables $\boldsymbol{s} \equiv \left[x_1, x_2,x_3\right]^\star$, occupied by
  the fluid at time $t$,
  \begin{equation}
  \nabla^p\, \eta \equiv\left\{\eta\,,_{ x_{j_1}} \ldots
  ,_{x_{j_p}}\right\} \quad \mathrm{and}\quad  \nabla ^{p}\rho
  \equiv \left\{\rho\,,_{ x_{j_1}} \ldots ,_{x_{j_p}}\right\}. \label{multigrad}
  \end{equation}
The subscript comma indicates   partial derivatives with respect to  variables  $
  x_{j_1} \ldots   x_{j_p} $ belonging to the set of Euler
  variables $( x_{1}, x_{2} , x_{3} ) $.  
 We deduce,
  \begin{equation*}
  d\varepsilon =\frac{\partial \varepsilon }{\partial \eta }\,d\eta +\frac{%
  	\partial \varepsilon }{\partial \rho }\,d\rho +\left(  \frac{\partial \varepsilon }{%
  	\partial \nabla \eta }\,\vdots\, d\nabla \eta\right) +\left(\frac{\partial \varepsilon }{%
  	\partial \nabla \rho }\,\vdots\, d\nabla \rho\right)   + \ldots +\left( \frac{\partial
  	\varepsilon }{\partial \nabla ^{n}\eta }\,\vdots\, d\nabla ^{n}\eta\right) +\left(\frac{%
  	\partial \varepsilon }{\partial \nabla ^{n}\rho }\,\vdots\, d\nabla
  ^{n}\rho\right) . \label{differentielenergy}
  \end{equation*}
Notation     \ $   \ \vdots\  
$\   means the complete product of tensors (or scalar
  product) and
  \begin{equation*}
  \tilde{T} =\frac{\partial \varepsilon (\rho,\eta)}{\partial \eta }\qquad \mathrm{and}%
  \qquad  \tilde{\mu} =\frac{\partial \varepsilon (\rho,\eta)}{\partial \rho },
  \end{equation*}%
 are  called the \emph{extended temperature} and \emph{extended chemical
  	potential}, respectively.

 \subsection{Equation of conservative motions} 
  The volume mass satisfies the mass
 conservation :
\begin{equation*}
 \frac{\partial \rho }{\partial t}+\rm{div}\left( \rho\, \boldsymbol{u}%
\right) =0 .
\label{density}
\end{equation*}
where 
$\boldsymbol{u}$ is the fluid velocity  and\ \ $\rm{div}$\ denotes the divergence operator. 
The motion is
supposed to be conservative and consequently, the volumetric entropy verifies :
 \begin{equation}
\frac{\partial \eta }{\partial t}+\rm{div}\left( \eta\, \boldsymbol{u}%
\right) =0 ,\label{entropyconservation}
\end{equation}
 The specific entropy $s=\eta /\rho$ is constant along each trajectory.
The \emph{extended divergence} \emph{at order }$p$ is defined as :
 \begin{equation*}
 {\rm{div}}_{p}(b_{{j_{1} \ldots  j_{p}}}) =\left( b_{{j_{1}\ldots  j_{p}}%
 }\right)_{{,x_{j_{1}}, \ldots   , x_{j_{p}}}},\  p\in \mathbb{N}\quad
 \mathrm{with}\quad {x_{j_{1}},\ldots , x_{j_{p}}}\in \left(
 x_{1}, x_{2} , x_{3}\right).
 \end{equation*}
 Classically, term $\left( b_{{j_{1}\ldots  j_{p}}%
 }\right)_{{,x_{j_{1}}, \ldots   ,x_{j_{p}}}}$ corresponds to the
 summation on the repeated indexes  $j_{1}\ldots  j_{p}$ of the
 consecutive derivatives of $ b_{j_{1}\ldots  j_{p}}$ with respect
 to $x_{j_{1}}, \ldots  ,  x_{j_{p}}$.
 Term ${%
 	\rm{div}}_{p}$ decreases from order $p$ the tensor order, while term $\nabla
 ^{p} $ increases from order $p$ the tensor order. 
 We denote :
 \begin{equation}
 \left\{
 \begin{array}{c}
 \displaystyle\theta =\tilde {T}-{\rm{div}\boldsymbol{\Psi }}_{\it 1}+{\rm{div}}_{\it 2}{%
 	\boldsymbol{\Psi }}_{\it 2}+{\ldots +}(-1)^{n}{\rm{div}}_{n}{\boldsymbol{\Psi }%
 }_{n}, \qquad {\rm with}\quad \boldsymbol{\Psi }_{p}=\frac{\partial \varepsilon }{\partial
 	\nabla ^{p}\eta }, \\
 \displaystyle\Xi =\tilde {\mu} -{\rm{div}  \boldsymbol{\Phi }}_{\it 1}+{\rm{div}}_{\it 2}{%
 	\boldsymbol{\Phi }}_{\it 2}+{\ldots +}(-1)^{n}{\rm{div}}_{n}{\boldsymbol{\Phi }%
 }_{n} ,\qquad {\rm with}\quad \boldsymbol{\Phi
 }_{p}=\frac{\partial \varepsilon }{\partial \nabla ^{p}\rho
 } ,
 \end{array}%
 \right.  \label{tempchem}
 \end{equation}%
  where  $\theta $ and $\Xi $\, are  called the
 \emph{generalized temperature} and \emph{generalized chemical
 	potential}. 
We obtain
the equation of conservative motions in Ref. \cite{Gouin}, where we can find the proofs of these results :
\begin{equation}
\boldsymbol{a}+ {\rm grad}\left( \Xi +\Omega \right)
+s\, {\rm grad}\,\theta = 0\qquad {\rm
	or}\qquad\boldsymbol{a}+ {\rm grad}\left( H+\Omega \right) -\theta \,
 {\rm grad}\, s=0 ,
\label{motion2}
\end{equation}%
where $\boldsymbol{a}$ denotes the acceleration, grad  the gradient operator, $\Omega$ the external force potential, $H=$ $\Xi -s\,\theta $  is called the    \emph{generalized free
	enthalpy}.
Relations \eqref{motion2} are the generalization of relation (29.8)
in Ref. \cite{Serrin} and constitutes the \emph{thermodynamic form} of the
equation of isentropic motions for perfect fluids.
\subsection{Complement: the stress tensor of conservative fluids}
 The new results of the subsection are not useful for the other parts of the paper, but completely extend  results obtained in \cite{casal4}.
We have the relation :
\begin{equation*}
d\varepsilon=\tilde T \ d\eta +\tilde\mu\ d\rho +\left({\boldsymbol{\Psi }}_{\it 1}\,\vdots\, d\nabla\eta\right)  +\left({ 
	\boldsymbol{\Phi }}_{\it 1}\,\vdots\,d\nabla \rho\right) + \ldots +\left(\boldsymbol{\Psi } 
_{n}\,\vdots\,   d\nabla ^{n}\eta\right) + \left(\boldsymbol{\Phi }_{n}\,\vdots\,d\nabla
^{n}\rho\right).
\end{equation*}
The Legendre
transformation of $\varepsilon$ with respect to $\ \eta, 
\rho, \nabla \eta, \nabla \rho, \dots,  \nabla ^{n}\eta,
\nabla ^{n}\rho$ is denoted by ${\it\Pi}$.  Fonction  ${\it\Pi}$
depends on
$\tilde T, \tilde\mu , {\boldsymbol{\Psi
}}_{\it 1}, {\boldsymbol{\Phi }}_{\it 1}, \dots,{\boldsymbol{\Psi }}_{n},
{\boldsymbol{\Phi }}_{n} $.
\begin{equation}
{\it\Pi} = \eta \ \tilde T + \rho \ \tilde\mu\  +
\left(\nabla \eta\,\vdots\, {\boldsymbol{\Psi }}_{\it 1}\right)  +\left(\nabla \rho\,\vdots\,{ 
	\boldsymbol{\Phi }}_{\it 1} \right) + \ldots +\left(\nabla ^{n}\eta\,\vdots\, \boldsymbol{\Psi } 
_{n}\right) + \left(\nabla
^{n}\rho\,\vdots\, \boldsymbol{\Phi }_{n}\right)
-\varepsilon,\label{Pi}
\end{equation}
and
\begin{equation*}
d{\it\Pi} = \eta\ d\tilde T \, +\ \rho\ d\tilde\mu\   +\left(
\nabla \eta\,\vdots\,d{\boldsymbol{\Psi }}_{\it 1}\right)+
\left(\nabla \rho \,\vdots\, d\boldsymbol{\Phi }_{\it 1}\right) +   \ldots +\left( \nabla ^{n}\eta \,\vdots\, d\boldsymbol{\Psi } 
_{n}\right)+ \left( \nabla ^{n}\rho\,\vdots\,d{\boldsymbol{\Phi }}_{n}\right),
\end{equation*}
where 
\begin{equation}
\frac{\partial{\it\Pi}}{\partial \tilde T}
=\eta,\quad\frac{\partial{\it\Pi}}{\partial \tilde\mu}
=\rho.\quad {\rm and}\ \quad \frac{\partial \mathcal{P}}{\partial  {\boldsymbol{\Psi }}_{\it k}}=  \nabla ^{k}\eta, \quad \frac{\partial \mathcal{P}}{\partial  {\boldsymbol{\Phi }}_{\it k}}=  \nabla ^{k}\rho ,\quad k \in \{\textit{1},	\ldots , n \} \label{pressure1}
\end{equation}
Consequently,%
\begin{equation*}
\frac{\partial {\it\Pi} }{\partial \boldsymbol{x}}=\ \eta \ \frac{\partial \tilde T}{%
	\partial \boldsymbol{x}}+\rho \frac{\partial \tilde\mu}{\partial \boldsymbol{x}}%
+\left( \nabla \eta \,\vdots\, \frac{\partial {\boldsymbol{\Psi }}_{\it 1}}{\partial
	\boldsymbol{x}}\right)  +  \left( \nabla \rho\,\vdots\,\frac{\partial {\boldsymbol{\Phi }}_{\it 1}}{%
	\partial \boldsymbol{x}}\right)+ \ldots + \left(\nabla ^{n}\eta \,\vdots\, \frac{\partial {%
		\boldsymbol{\Psi }}_{n}}{\partial \boldsymbol{x}}\right)+\left(\nabla ^{n}\rho \,\vdots\, \frac{%
	\partial {\boldsymbol{\Phi }}_{n}}{\partial \boldsymbol{x}}\right) .
\end{equation*}
Because $\quad \displaystyle \frac{\partial \,{\rm{div}}_{n}{\,\boldsymbol{\Phi }_n}%
}{\partial \boldsymbol{x}} =
{\rm{div}}_{n} \frac{\,\partial\boldsymbol{\Phi}_{n}
} {\partial\boldsymbol{x}} $, \ by taking  account of identities,
\begin{eqnarray}
\left(\nabla \rho \,\vdots\, \frac{\partial {\boldsymbol{\Phi }}_{\it 1}}{\partial
	\boldsymbol{x}}\right) &\equiv&{\rm{div}}\left(\, \rho \ \frac{\partial {\boldsymbol{%
			\Phi }}_{\it 1}}{\partial \boldsymbol{x}}\right) -\rho \ \frac{\partial \,{\rm{%
			div}\boldsymbol{\Phi }}_{1}}{\partial \boldsymbol{x}}  \notag \\
\left(\nabla ^{\it 2}\rho \,\vdots\, \frac{\partial {\boldsymbol{\Phi }}_{\it 2}}{\partial
	\boldsymbol{x}}\right) &\equiv&{\rm{div}}\left[\left( \nabla \rho \,\vdots\, \ \frac{\partial {%
		\boldsymbol{\Phi }}_{\it 2}}{\partial \boldsymbol{x}}\right)-\rho \ \frac{\partial {%
		\rm{div}\ \boldsymbol{\Phi }}_{\it 2}}{\partial \boldsymbol{x}}\right] +\rho \
\frac{\partial \,{\rm{div}}_{\it 2}{\,\boldsymbol{\Phi }}_{\it 2}}{\partial
	\boldsymbol{x}}  \notag \\
&\vdots &  \label{lemme1} \\
\left(\nabla ^{n}\rho \,\vdots\,\frac{\partial {\boldsymbol{\Phi }}_{n}}{\partial
	\boldsymbol{x}}\right) &\equiv&{\rm{div}}\left[\left( \nabla ^{n-1}\rho \,\vdots\,   \frac{%
	\partial {\boldsymbol{\Phi }}_{n}}{\partial \boldsymbol{x}}\right)-\left(\nabla
^{n-2}\rho \,\vdots\, \rm{div}\frac{\partial {\boldsymbol{\Phi }}_{\it n}}{%
	\partial \boldsymbol{x}}\right)+{\ldots }\right. \   \notag \\
&+& \left. (-1)^{p-1}\left(\nabla ^{n-p}\rho \,\vdots\,{\rm{div}}_{p-1}\frac{%
	\partial {\boldsymbol{\Phi }}_{n}}{\partial \boldsymbol{x}}\right)+{\ldots +}%
(-1)^{n-1}\rho\, \,{\rm{div}}_{n-1}\frac{\partial {\boldsymbol{\Phi }}_{n}%
}{\partial \boldsymbol{x}}\right]   
+(-1)^{n}\rho \ \frac{\partial \,{\rm{div}}_{n}{\,\boldsymbol{\Phi }}_{n}%
}{\partial \boldsymbol{x}},  \notag
\end{eqnarray}%
and  an analog expression for $\eta $, where $\rho$ and $ {\boldsymbol{\Phi }}_{p}$ are replaced by $\eta$ and $ {\boldsymbol{\Psi }}_{p}$, $p\in \{1,\ldots ,n\}$. We deduce,
\begin{equation*}
\rho \frac{\partial \Xi }{\partial \boldsymbol{x}}+\eta \frac{\partial
	\theta }{\partial \boldsymbol{x}}=\rm{div}\left( \,{\it\Pi} \, \boldsymbol{I}-%
\boldsymbol{\sigma }\,\right) .
\end{equation*}
The identical transformation is denoted by
$\boldsymbol{I}$ and   stress tensor $\sigma$ is :
\begin{eqnarray*}
	\boldsymbol{\sigma } &=& {\rho \ }\frac{\partial {\boldsymbol{\Phi
		}}_{\it 1}}{\partial \boldsymbol{x}}+\left(\nabla \rho \,\vdots\, \frac{\partial {%
			\boldsymbol{\Phi }}_{\it 2}}{\partial \boldsymbol{x}}\right)-\rho \ \frac{\partial \,{%
			\rm{div}}_{\it 2}{\,\boldsymbol{\Phi }}_{\it 2}}{\partial \boldsymbol{x}}+ \ldots + 
	\left( \nabla ^{n-1}\rho  \,\vdots\,\frac{\partial {\boldsymbol{\Phi
		}}_{n}}{\partial
		\boldsymbol{x}}\right)-\left(\nabla ^{n-2}\rho   \,\vdots\,  \rm{div}\frac{\partial {%
			\boldsymbol{\Phi }}_{\it n}}{\partial \boldsymbol{x}}\right)     
	\\
	&+&	\ldots +  (-1)^{p-1}\left(\nabla ^{n-p}\rho \,\vdots\,{\rm{div}}_{p-1}\frac{%
		\partial {\boldsymbol{\Phi }}_{n}}{\partial \boldsymbol{x}} \right)+  \ldots +(-1)^{n-1}\rho \,{\rm{div}}_{n-1}\frac{\partial {\boldsymbol{\Phi }}_{n}}{%
		\partial \boldsymbol{x}}   \\
	&+&{\eta\ }\frac{\partial {\boldsymbol{\Psi
		}}_{\it 1}}{\partial \boldsymbol{x}}+\left(\nabla \eta\,\vdots\, \frac{\partial {%
			\boldsymbol{\Psi }}_{\it 2}}{\partial \boldsymbol{x}}\right)-\eta\ \frac{\partial \,{%
			\rm{div}}_{\it 2}{\,\boldsymbol{\Psi }}_{\it 2}}{\partial \boldsymbol{x}}+ \ldots + 
	\left( \nabla ^{n-1}\eta \,\vdots\,\frac{\partial {\boldsymbol{\Psi
		}}_{\it n}}{\partial
		\boldsymbol{x}}\right)-\left(\nabla ^{n-2}\eta  \,\vdots\,  \rm{div}\frac{\partial {{\boldsymbol{\Psi }}}_{\it n}}{\partial \boldsymbol{x}}\right)   
	\\
	& + &\ldots + 	 (-1)^{p-1}\left(\nabla ^{n-p}\eta\,\vdots\,{\rm{div}}_{p-1}\frac{%
		\partial {\boldsymbol{\Psi }}_{n}}{\partial \boldsymbol{x}} \right)+  \ldots +(-1)^{n-1}\eta\,{\rm{div}}_{n-1}\frac{\partial {\boldsymbol{\Psi }}_{n}}{%
		\partial \boldsymbol{x}} .
\end{eqnarray*}
Due to the mass conservation, we get
$\
\rho \,\boldsymbol{a} = {\partial (\rho\,\boldsymbol{u})}/{\partial
	t}+\rm{div}\left( \rho
\boldsymbol{u}\otimes \boldsymbol{u}\right) \
$
and the
equation of motions \eqref{motion2} can be written in the other form :
\begin{equation}
\frac{\partial \rho \,\boldsymbol{u}}{\partial t}+\rm{div}\left(\, \rho
\boldsymbol{u}\otimes \boldsymbol{u}+ \,{\it\Pi} \, \boldsymbol{I}-\boldsymbol{%
	\sigma }\,\right) + \rho \, \rm{grad}\, \Omega=0\label{stress tensor}
\end{equation}
The two previous equations are deduced from Hamilton's principle \cite{Gouin}, which can be used only for conservative media because  
Eq. \eqref{entropyconservation} is verified. In this case,
Eq. \eqref{motion2} is strictly equivalent to Eq. \eqref{stress tensor}.\\ 
Let us note that, for classical fluids, the two equations are two  forms of the equation of motions which are written in Eq. (29.8) of \cite{Serrin}: 
\begin{equation*}
\rho \,\boldsymbol{a}+ {\rm grad}\, p   +\rho\,{\rm grad} \, \Omega  
=0 \qquad {\Longleftrightarrow
	 }\qquad  \boldsymbol{a}+ {\rm grad}\left( \mu+\Omega \right)
 +s\, {\rm grad}\,T = 0,
\label{motion}
\end{equation*}
where  $p$ is here the thermodynamical pressure of simple fluids, $\mu$ the corresponding chemical potential and  $T$ the Kelvin temperature.\\
The stress tensor $\boldsymbol{%
	\sigma }$ is only an artifact different from the Cauchy stress tensor, which can be interesting to compare with solid mechanics; the most important conservative equations are expressed by Eq. \eqref{motion2}. It is the reason why the entropy law is  expressed without dissipative terms.

\subsection{Equation of energy for dissipative motions  (see  the detailed proofs in Ref. \cite{Gouin})}

For  viscous
fluids, the equation of motions
 can be written as :
\begin{equation*}
\frac{\partial \rho \,\boldsymbol{u}}{\partial t}+\rm{div}\left(
\rho\boldsymbol{u}\otimes \boldsymbol{u}\right)+\rho \,\rm{grad}\, \Xi +\eta
\, \rm{grad}\, \theta -\rm{div}\,{\boldsymbol{\sigma}}_{v} + \rho
\, \rm{grad}\, \Omega =0 ,
\end{equation*}
here $\boldsymbol{\sigma }_{v}$ denotes the viscous-stress tensor
of the fluid. We denote
\begin{equation*}
\left\{
\begin{array}{l}
\displaystyle\boldsymbol{M}  =  \frac{\partial  \rho
\boldsymbol{u}}{\partial t}+\rm{div}\left( \rho
\boldsymbol{u}\otimes \boldsymbol{u}\right) +\rho \,\rm{grad}\, {%
	\Xi } +\eta \,\rm{grad} \, \theta -\rm{div}\boldsymbol{\sigma
}_{\it v}+\rho
\, \rm{grad}\, \Omega  \\
\displaystyle B  = \frac{\partial \rho }{\partial t}+\rm{div}\left( \rho \,%
\boldsymbol{u}\right) \\
\displaystyle N  = \frac{\partial\rho s}{\partial t}+{\rm{div}}\left(\rho s\, \boldsymbol{u} \right)
+\frac{1}{\theta}\,\Big(\rm{div}{\boldsymbol q}
-r-\rm{Tr}\,\big(\,
\boldsymbol{\sigma }_{\it v}\, \boldsymbol{D }\, \big)\Big)
\\
\displaystyle F  =  \frac{\partial }{\partial t}\left(
\frac{1}{2}\,\,\rho\, \boldsymbol{u%
}\, \boldsymbol{.}\,\boldsymbol{u}+\rho \ \Xi
+\eta \ \theta -{\it\Pi} +\rho \ \Omega \right)\\
\displaystyle+\rm{div}\left\{ \left[ \left( \frac{1}{2}\,\rho\, \boldsymbol{u%
}\, \boldsymbol{.}\,\boldsymbol{u}+\rho \,\Xi +\eta \,\theta +\rho \,\Omega
\right)
\boldsymbol{I} - \boldsymbol{\sigma }_{\it v}\right] \boldsymbol{u}+\boldsymbol{\chi }%
\right\} +\rm{div}\,\boldsymbol{q}-r-\rho\,\frac{\partial \Omega
}{\partial t},
\end{array}%
\right.
\end{equation*}
where \ Tr\ \ denotes the trace operator and\ \  $  \boldsymbol{.} $ \ the scalar product ($\boldsymbol{u}\,\boldsymbol{.}\,\boldsymbol{u} =\boldsymbol{u}^\star \boldsymbol{u}$). Terms $\boldsymbol{q}$ and  $r$ represent the heat-flux vector and the heat supply;   $D=\frac{1}{2}\left( \partial\boldsymbol{u}/\partial \boldsymbol{x}+(\partial\boldsymbol{u}/\partial \boldsymbol{x})^\star\right) $ is the velocity gradient. Due to the relaxation time in the dissipative processes, we  only consider the case when  dissipative viscous stress tensor $\boldsymbol{\sigma
}_{\it v}$ takes account of the first derivative of the velocity field: the higher   terms are assumed negligible and   the viscosity  does not take any gradient terms into
account.\\
The equation of motion is written $M=0$   in the    dissipative case with addition of   viscous stress tensor   $\boldsymbol{\sigma
}_{\it v}$.     The equation of motion
	written in the conservative case can be now written for  viscous
	fluids,  the  conservative motions are written  without viscosity. Terms $\boldsymbol{q}$ and $r$ being  introduced together with $\boldsymbol{\sigma
}_{\it v}$ are adapted into   $N$ and $F$ for the dissipative case.\\
 Due to subsection 2.3, the term $\boldsymbol{M} $ can be written in two equivalent expressions :
\begin{equation*}
\boldsymbol{M}  =  \frac{\partial  \rho
	\boldsymbol{u}}{\partial t}+\rm{div}\left( \rho
\boldsymbol{u}\otimes \boldsymbol{u}\right) +\rho \,\rm{grad}\, {%
	\Xi } +\eta \,\rm{grad} \, \theta -\rm{div}\boldsymbol{\sigma
}_{\it v}+\rho
\, \rm{grad}\, \Omega
\end{equation*}
or equivalently,
\begin{equation*}
\boldsymbol{M}=\frac{\partial \rho \,\boldsymbol{u}}{\partial t}+\rm{div}\left(\, \rho
\boldsymbol{u}\otimes \boldsymbol{u}+ \,{\it\Pi} \, \boldsymbol{I}-\boldsymbol{%
	\sigma} - \boldsymbol{\sigma
}_{\it v} \,\right) + \rho \, \rm{grad}\, \Omega 
\end{equation*}
The first expression is more adapted to the following.
Viscous stress  tensor  $\boldsymbol{\sigma
}_{\it v}$ is classically introduced in the same part than conservative stress tensor $\boldsymbol{\sigma
}$.  In this case,
\begin{equation*}
N  = \frac{\partial\rho s}{\partial t}+{\rm{div}}\left(\rho s\, \boldsymbol{u} \right)
+\frac{1}{\theta}\,\Big(\rm{div}{\boldsymbol q}
-r-\rm{Tr}\,\big(\,
\boldsymbol{\sigma }_{\it v}\, \boldsymbol{D }\, \big)\Big)
\end{equation*}
  and due to relation,
\begin{equation*}
\left( \frac{\partial\rho s}{\partial t} +{\rm{div}}\left(\rho s\, \boldsymbol{u} \right)\right) {\theta}
+ \,\Big(\rm{div}{\boldsymbol q}
-r-\rm{Tr}\,\big(\,
\boldsymbol{\sigma }_{\it v}\, \boldsymbol{D }\, \big)\Big)=0,
\end{equation*}
term $ \displaystyle  \frac{\partial\rho s}{\partial t}+{\rm{div}}\left(\rho s\, \boldsymbol{u} \right)$ corresponds to the variation of the entropy. 
    Then, we only take account  of 
 $ \boldsymbol{D}$  which is  the velocity deformation tensor and (see for proof, Ref. \cite{Gouin}),
\begin{eqnarray*}
	\boldsymbol{\chi} = \rho \frac{\partial \Phi _{\it 1}}{\partial t}+\ldots+\left(\frac{\partial \Phi
		_{n}}{\partial t}
	\,\vdots\, \nabla ^{n-1}\rho\right)-\left( \rm{div}\frac{\partial \Phi _{ \it n}}{%
		\partial t}
	\,\vdots\, \nabla ^{{\it n}-2}\rho 
	 \right)+\ldots +(-1)^{p-1}\left( {\rm{div}}_{p-1}\frac{\partial
	 	\Phi
	 	_{n}}{\partial t}
	  \,\vdots\,
	 \nabla ^{n-p}\rho \right)
	+\ldots  +(-1)^{n-1}\,\rho \, {\rm{div}}_{n-1}\frac{%
		\partial \Phi _{n}}{\partial t} \\
	+\,\eta
	\,\frac{\partial \Psi _{\it 1}}{\partial t}+ \ldots  	 +
	  \left(\frac{\partial \Psi
	  	_{\it n}}{\partial \it t}
	\,\vdots\, 
	\nabla ^{n-1}\eta
	\right)
	-	\left(\rm{div}\frac{\partial \Psi _{\it n}}{%
		\partial \it t}\,\vdots\,  \nabla ^{{\it n}-2}\eta\right)
	+{\ldots }+(-1)^{p-1}
	\left({\rm{div}}_{p-1}\frac{\partial
		\Psi
		_{\it n}}{\partial \it t} \,\vdots\, 
	\nabla ^{n-p}\eta \right)
	+\ldots  +(-1)^{n-1}\,\eta \,{\rm{div}}_{n-1}\frac{%
		\partial \Psi _{n}}{\partial t} .
\end{eqnarray*}
Term $\boldsymbol{\chi}$ is the general extension of the interstitial-working vector
   obtained in \cite{casal5}.
 We obtain the following results (see  the  proofs in Ref. \cite{Gouin}),
\begin{theorem} 
Relation%
\begin{equation*}
F-\boldsymbol{M}\, \boldsymbol{.}\,\boldsymbol{u}-\left( \frac{1}{2}\,\boldsymbol{u}\, \boldsymbol{.}\,
 \boldsymbol{u}+\ \Xi +\ \Omega \right) \,B-\theta\,N\equiv
0
\end{equation*}%
is an algebraic identity.
\end{theorem} 	
\noindent	$\boldsymbol{M} = 0$\ is the equation of motion, $B=0$ is the mass
	conservation and $N=0$ the entropy relation, then $F=0$ is the
	equation of energy for dissipative fluids.  
	\begin{corollary}
	The equation of energy is
	\begin{equation*}\label{eenrgy}
	\displaystyle\frac{\partial }{\partial t}\left( \frac{1}{2}\,\rho\,
	\boldsymbol{u}\, \boldsymbol{.}\,
	\boldsymbol{u}
	+\rho \ \Xi +\eta \ \theta -{\it\Pi}+\rho \ \Omega \right) + 
	\displaystyle\rm{div}\left\{ \left[ \left( {\frac{1}{2}}\,\rho\,\boldsymbol{u}\, \boldsymbol{.}\,
	\boldsymbol{u}+\rho \,\Xi +\eta \,\theta +\rho \,\Omega
	\right)
	\boldsymbol{I}-\boldsymbol{\sigma }_{\it v}\right] \boldsymbol{u}+\boldsymbol{\chi }%
	\right\} +\rm{div}\, \boldsymbol{q}-r-\rho\,\frac{\partial \Omega }{\partial t}=0.
	\end{equation*}
\end{corollary}
For  dissipative fluid motions,
$
tr\left( \boldsymbol{\sigma }_{v}\,D\right)\geq 0
$.
From $N = 0$ and $B=0$, we deduce the Planck inequality
\cite{Truesdel1} :
\begin{equation*}
\rho \,\theta \,\frac{ds}{dt}+\rm{div}\,\boldsymbol{q} - r \geq 0.
\end{equation*}
We consider the Fourier equation  in the   form of general inequality :
$$
\boldsymbol{q}\,\boldsymbol{.}\,\rm{grad}\, \theta \leq 0 ,
$$ 
and we obtain,
\begin{equation*}
\rho \,\frac{ds}{dt}+\rm{div}\,\frac{\boldsymbol{q}}{\theta }-\frac{r}{%
	\theta } \geq 0,
\end{equation*}
which is the extended form of the Clausius-Duhem inequality. Then, multi-gradient fluids are compatible with the two law of thermodynamics.

 \section{Main field variables}

In this section, we use the  properties of symmetry and consequently we cannot any more identify covariant and contravariant vectors and tensors. Then, superscript\ $^{\star}$\ denotes the
transposition in ${\mathcal D}_t$.   When clarity is necessary,   we  use    the notation $\boldsymbol{b%
}^{\star }\boldsymbol{c}$ for the scalar product  of vectors  $\boldsymbol{b}$ and $\boldsymbol{c}$,
tensor product  $\boldsymbol{b}\, \boldsymbol{c}^{\star }$ corresponds to
$\boldsymbol{b}\otimes \boldsymbol{c}$.  
The divergence of a linear
transformation $\boldsymbol{S}$ denotes the covector $\mathop{\rm div}(%
\boldsymbol{S})$ such that, for any constant vector $\boldsymbol{d}$, $%
\mathop{\rm div}(\boldsymbol{S})\, \boldsymbol{d}= \mathop{\rm div}(%
\boldsymbol{S}\,\boldsymbol{d})$.  
Now, previous terms $\nabla ^{p}\eta$ and $\nabla ^{p}\rho$, defined in  Eqs. (\ref{multigrad}),   are covariant tensors of order $p$, while $\boldsymbol{\Psi
}_{p}$ and $\boldsymbol{\Phi
}_{p}$, defined in Eqs. \eqref{tempchem}, are contravariant  tensors of order $p$.

 \subsection{Study of   conservative motion equation}

 Without missing the generality, and for the sake of simplicity, we do not consider external-force term. The total energy of the fluid is :
 \begin{equation*}
 E=\frac{\boldsymbol{j}^{\star}\boldsymbol{j}}{2\rho }+\varepsilon
 \qquad \mathrm{where}\quad \boldsymbol{j}=\rho \, \boldsymbol{u},
 \end{equation*}%
 and
 \begin{equation*}
 dE=\tilde T \ d\eta +R\ d\rho +\boldsymbol{u}^\star d\boldsymbol{j}+\left( d\nabla \eta\,\vdots\, {\boldsymbol{\Psi }}_{\it {\LARGE 1}}\right) +\left(d\nabla \rho\,\vdots\,   {%
 	\boldsymbol{\Phi }}_{\it 1}\right) +\ldots +\left(d\nabla ^{n}\eta \,\vdots\,   {\boldsymbol{\Psi }}%
 _{n}\right) +\left(d\nabla ^{n}\rho \,\vdots\,    {\boldsymbol{\Phi
 }}_{n}\right) ,
 \end{equation*}
 where
 \begin{equation*}
 R=\tilde\mu \ -%
 \frac{\boldsymbol{u}^{\star }\boldsymbol{u}}{2}.
 \end{equation*}
 The Legendre
 transformation of $E$ with respect to variables $\, \eta, \,
 \rho, \boldsymbol{j}, \nabla \eta, \nabla \rho,\dots, \nabla
 ^{n}\eta, \nabla ^{n}\rho$ is denoted $ \mathcal{P} $; $ \mathcal{P} $  is a function
 of
 $\tilde T, R, \boldsymbol{u}, {\boldsymbol{\Psi
 }}_{\it 1}, {\boldsymbol{\Phi }}_{\it 1}, \dots, {\boldsymbol{\Psi }}_{n},
 {\boldsymbol{\Phi }}_{n} $.
 \begin{equation*}
 \mathcal{P}=  \eta \ \tilde T+\ \rho \ R +\boldsymbol{j}^\star  \boldsymbol{u}+\left(\nabla \eta\,\vdots\, {\boldsymbol{\Psi }}_{\it 1}\right) +\left(\nabla \rho
 \,	\vdots\, {\boldsymbol{\Phi }}_{\it 1}\right)+ \ldots  +\left(\nabla^n \eta\,\vdots\, {\boldsymbol{\Psi }}_{n}\right)+\left(\nabla^n \rho
 \,	\vdots\, {\boldsymbol{\Phi }}_{n}\right) -E,\label{pressure0}
 \end{equation*}
 where
 \begin{equation}
 \frac{\partial \mathcal{P}}{\partial \tilde T}
 =\eta,\quad\frac{\partial \mathcal{P}}{\partial R}
 =\rho,\quad\frac{\partial\mathcal{P}}{\partial \boldsymbol{u}}
 =\boldsymbol{j}\,^\star,\quad {\rm and}\ \quad \frac{\partial \mathcal{P}}{\partial  {\boldsymbol{\Psi }}_{\it k}}=  \nabla ^{k}\eta, \quad \frac{\partial \mathcal{P}}{\partial  {\boldsymbol{\Phi }}_{\it k}}=  \nabla ^{k}\rho ,\quad k \in \{\textit{1},	\ldots , n \}\label{pressure1}.
 \end{equation}
We notice that
 \begin{equation}
\mathcal{P}=  \eta \ \tilde T+\ \rho \  \tilde\mu  +\left(\nabla \eta\,\vdots\, {\boldsymbol{\Psi }}_{\it 1}\right) +\left(\nabla \rho
\,	\vdots\, {\boldsymbol{\Phi }}_{\it 1}\right)+ \ldots  +\left(\nabla^n \eta\,\vdots\, {\boldsymbol{\Psi }}_{n}\right)+\left(\nabla^n \rho
\,	\vdots\, {\boldsymbol{\Phi }}_{n}\right) -\ \varepsilon.\label{pressure}
\end{equation}
Consequently, the value of  $\mathcal{P}$ is the same than the value of $\it{\Pi} $ given in Eq. \eqref{Pi},  but function  $\mathcal{P}$ is associated with a different field of variables. 
Motion equation \eqref{motion2} can be written,
 \begin{equation}
 \frac{\partial \rho \,\boldsymbol{u}}{\partial t}+\rm{div}\left( \rho
 \boldsymbol{u}\otimes \boldsymbol{u} \right) +\rho \,\rm{grad}\,\Xi +\eta \,\rm{grad}%
 \,\theta = 0 .  \label{motion1}
 \end{equation}%
 Due to \eqref{pressure1},
 \begin{equation*}  
 d\left( \mathcal{P} \,\boldsymbol{u}\right)  = d\mathcal{P} \ \,\boldsymbol{u}+\mathcal{P} \ d%
 \boldsymbol{u}\quad \mathrm{\Longrightarrow }\quad \frac{\partial
 	\left( \mathcal{P}
 	\,\boldsymbol{u}\right) }{\partial \boldsymbol{u}}=\boldsymbol{u}\otimes \boldsymbol{j} +
 \mathcal{P} \ \boldsymbol{I}  \notag \quad
 \mathrm{\Longrightarrow } \quad \rm{div}\left[ \frac{\partial
 	\left( \mathcal{P} \,\boldsymbol{u}\right) }{\partial
 	\boldsymbol{u}}\right] = \rm{div}\left( \rho \boldsymbol{u\otimes
 	u}\right)  +\frac{\partial \mathcal{P} }{\partial \boldsymbol{x}}.
 \end{equation*}
 From \eqref{pressure}, we get
 \begin{equation*}
 	\frac{\partial \mathcal{P} }{\partial \boldsymbol{x}}=\eta \ \frac{\partial \tilde T}{%
 		\partial \boldsymbol{x}}+\rho \ \frac{\partial \tilde\mu }{\partial \boldsymbol{x}}%
 	\ +\left(\nabla \eta\, \vdots\, \frac{\partial {\boldsymbol{\Psi }}_{\it 1}}{\partial
 		\boldsymbol{x}}\right)+\left(\nabla \rho \, \vdots\, \frac{\partial \boldsymbol{\Phi }_{\it 1}}{%
 		\partial \boldsymbol{x}}\right)+ \ldots + \left(\nabla ^{n}\eta \, \vdots\, \frac{\partial {%
 			\boldsymbol{\Psi }}_{n}}{\partial \boldsymbol{x}}\right)+\left(\nabla ^{n}\rho\, \vdots\, \frac{%
 		\partial {\boldsymbol{\Phi }}_{n}}{\partial \boldsymbol{x}}\right) .
 \end{equation*}%
 Taking account of Eq. \eqref{lemme1}, we get :
 \begin{eqnarray}
 	\frac{\partial \mathcal{P} }{\partial \boldsymbol{x}} &=&\eta \ \frac{\partial }{\partial \boldsymbol{x}}\Big(\, \tilde T-{\rm{div}\,
 		\boldsymbol{\Psi }}_{\it 1}+{\rm{div}}_{\it 2}{\boldsymbol{\Psi }}_{\it 2}+ \ldots +
 	(-1)^{n}{\rm{div}}_{n}{\boldsymbol{\Psi }}_{n}\,\Big) 
 + \rho \ \frac{\partial }{%
 		\partial \boldsymbol{x}}\Big(\, \tilde\mu -{\rm{div}\ \boldsymbol{\Phi }}_{\it 1}+{%
 		\rm{div}}_{\it 2}{\boldsymbol{\Phi }}_{\it 2}+{\ldots +}(-1)^{n}{\rm{div}}_{n}{%
 		\boldsymbol{\Phi }}_{n}\,\Big)\notag \\
 	&+& {\rm{div}}\left( \eta \ \frac{\partial {\boldsymbol{\Psi }}_{\it 1}}{%
 		\partial \boldsymbol{x}}\right) +{\rm{div}}\left[ \left( \nabla \eta \,\vdots\, \frac{%
 		\partial {\boldsymbol{\Psi }}_{\it 2}}{\partial \boldsymbol{x}}\right)-\eta \ \frac{%
 		\partial {\rm{div}\ \boldsymbol{\Psi }}_{\it 2}}{\partial \boldsymbol{x}}%
 	\right]\notag\\
 	& \vdots &\notag \\
 	&+&{\rm{div}}\left[ \left(\nabla ^{n-1}\eta \,\vdots\,   \frac{\partial {\boldsymbol{%
 				\Psi }}_{\it n}}{\partial \boldsymbol{x}}\right)-\left(\nabla ^{n-2}\eta \,\vdots\, \rm{div}%
 	\frac{\partial {\boldsymbol{\Psi }}_{\it n}}{\partial \boldsymbol{x}}\right)+{\ldots }%
 +(-1)^{p-1}\left(\nabla ^{n-p}\eta \,\vdots\, {\rm{div}}_{p-1}\frac{%
 		\partial {\boldsymbol{\Psi }}_{n}}{\partial \boldsymbol{x}}\right)+ \ldots + 
 	(-1)^{n-1}\eta \,\,{\rm{div}}_{n-1}\frac{\partial {\boldsymbol{\Psi }}_{n}%
 	}{\partial \boldsymbol{x}}\right]\notag\\
 	&+&{\rm{div}}\left( \rho \ \frac{\partial {\boldsymbol{\Phi }}_{\it 1}}{%
 		\partial \boldsymbol{x}}\right) +{\rm{div}}\left[ \left(\nabla \rho \,\vdots\, \frac{%
 		\partial {\boldsymbol{\Phi }}_{\it 2}}{\partial \boldsymbol{x}}\right)-\rho \ \frac{%
 		\partial {\rm{div}\ \boldsymbol{\Phi }}_{\it 2}}{\partial \boldsymbol{x}}%
 	\right] \label{Key1}\\
 	& \vdots& \notag\\
 	&+&{\rm{div}}\left[\left(\nabla ^{n-1}\rho\,\vdots\,   \frac{\partial {\boldsymbol{%
 				\Phi }}_{\it n}}{\partial \boldsymbol{x}}\right)-\left(\nabla ^{n-2}\rho \,\vdots\, \rm{div}%
 	\frac{\partial {\boldsymbol{\Phi }}_{\it n}}{\partial \boldsymbol{x}}\right)+ \ldots 
 	 +  (-1)^{p-1}\left(\nabla ^{n-p}\rho \,\vdots\,{\rm{div}}_{p-1}\frac{%
 		\partial {\boldsymbol{\Phi }}_{n}}{\partial \boldsymbol{x}}\right)+{\ldots +}%
 	(-1)^{n-1}\rho \,\,{\rm{div}}_{n-1}\frac{\partial {\boldsymbol{\Phi }}_{n}%
 	}{\partial \boldsymbol{x}}\right]\notag
 \end{eqnarray}%
 and consequently,
 \begin{eqnarray*}
 	\frac{\partial \mathcal{P} }{\partial \boldsymbol{x}} &=&\eta \
 	\frac{\partial \theta }{\partial \boldsymbol{x}}+\rho \
 	\frac{\partial \Xi }{\partial
 		\boldsymbol{x}} +\rm{div}\left(\, C_{\it n}+D_{\it n}\,\right)
   \\
 &+& \rm{div}\left[\, \eta \, \frac{\partial }{\partial
 	\boldsymbol{x}}\left(
 {\boldsymbol{\Psi }}_{\it 1}-{\rm{div}}{\boldsymbol{\Psi}}%
 _{\it 2}+{\ldots +}(-1)^{{\it n}-1}{\rm{div}}_{{\it n}-1}{\boldsymbol{\Psi}}_{\it n}\right) %
 \right] +\rm{div}\left[ \,\rho \, \frac{\partial }{\partial \boldsymbol{x}}\left(
 	{\boldsymbol{\Phi }}_{\it 1}-{\rm{div}}{\boldsymbol{\Phi }%
 	}_{\it 2}+{\ldots +}(-1)^{{\it n}-1}{\rm{div}}_{{\it n}-1}{\boldsymbol{\Phi}}_{\it n}\right) %
 	\right]   ,
 \end{eqnarray*}%
 with%
 \begin{eqnarray}
 	C_{n} &=&\left(\nabla \eta\,\vdots\, \frac{\partial {\boldsymbol{\Psi }}_{\it 2}}{%
 		\partial \boldsymbol{x}}\right)+\left(\nabla ^{\it 2}\eta \,\vdots\, \frac{\partial {\boldsymbol{%
 				\Psi }}_{\it 3}}{\partial \boldsymbol{x}}\right)-\left(\nabla \eta \,\vdots\, \rm{div}\frac{%
 		\partial {\boldsymbol{\Psi }}_{\it 3}}{\partial \boldsymbol{x}}\right)+ \ldots  
+\left(\nabla ^{n-1}\eta\,\vdots\, \frac{\partial {\boldsymbol{\Psi }}_{n}}{%
 		\partial \boldsymbol{x}}\right)-\left(\nabla ^{n-2}\eta \,\vdots\,\rm{div}\frac{\partial {%
 			\boldsymbol{\Psi }}_{\it n}}{\partial \boldsymbol{x}}\right)+ \ldots\label{Key2} \\
 	&+& (-1)^{p-1}\left(\nabla ^{n-p}\eta\,\vdots\,{\rm{div}}_{p-1}\frac{\partial {%
 			\boldsymbol{\Psi }}_{n}}{\partial \boldsymbol{x}}\right)+\ldots +(-1)^{n-2}\left(\nabla
 	\eta \,\vdots\,{\rm{div}}_{n-2}\frac{\partial {\boldsymbol{\Psi }}_{n}}{%
 		\partial \boldsymbol{x}}\right),\notag
\\
 	D_{n} &=&\left(\nabla \rho\,\vdots\, \frac{\partial {\boldsymbol{\Psi }}_{\it 2}}{%
 		\partial \boldsymbol{x}}\right)+\left(\nabla ^{\it 2}\rho \,\vdots\, \frac{\partial {\boldsymbol{%
 				\Psi }}_{\it 3}}{\partial \boldsymbol{x}}\right)-\left(\nabla \rho \,\vdots\, \rm{div}\frac{%
 		\partial {\boldsymbol{\Psi }}_{\it 3}}{\partial \boldsymbol{x}}\right)+ \ldots  
 	+\left(\nabla ^{n-1}\rho\,\vdots\, \frac{\partial {\boldsymbol{\Psi }}_{n}}{%
 		\partial \boldsymbol{x}}\right)-\left(\nabla ^{n-2}\rho \,\vdots\,\rm{div}\frac{\partial {%
 			\boldsymbol{\Psi }}_{n}}{\partial \boldsymbol{x}}\right)+ \ldots \label{Key3}\\
 	&+& (-1)^{p-1}\left(\nabla ^{n-p}\rho\,\vdots\,{\rm{div}}_{p-1}\frac{\partial {%
 			\boldsymbol{\Psi }}_{n}}{\partial \boldsymbol{x}}\right)+\ldots +(-1)^{n-2}\left(\nabla
 	\rho \,\vdots\,{\rm{div}}_{n-2}\frac{\partial {\boldsymbol{\Psi }}_{n}}{%
 		\partial \boldsymbol{x}}\right) .\notag
 \end{eqnarray}%
From Eq. \eqref{motion1} and Eqs. \eqref{Key1}-\eqref{Key2}-\eqref{Key3}, we
 finally obtain,
 \begin{eqnarray*}
 	&&\frac{\partial \rho \,\boldsymbol{u}}{\partial t} +\rm{div}\left[ \frac{%
 		\partial \left( \mathcal{P} \,\boldsymbol{u}\right) }{\partial \boldsymbol{u}}\right]
 -\rm{div}\left( \, C_{\it n}+D_{\it n}\,\right) 	\\
 	&& -\,\rm{div}\left[ \eta \,\frac{\partial }{\partial
 		\boldsymbol{x}}\left(
 	\boldsymbol{\Psi}_{\it 1}-{\rm{div}} {\boldsymbol{\Psi}}_{\it 2}+%
 	\ldots + (-1)^{{\it n}-1}{\rm{div}}_{{\it n}-1}{\boldsymbol{\Psi }}_{\it n}\right) \right]  
 	 -  \rm{div}\left[ \rho \, \frac{\partial }{\partial \boldsymbol{x}}\left(
 	\boldsymbol{\Phi}_{\it 1}-{\rm{div}}{\boldsymbol{\Phi }%
 	}_{\it 2}+{\ldots }+(-1)^{{\it n}-1}{\rm{div}}_{{\it n}-1}{\boldsymbol{\Phi}}_{\it n}\right) %
 	\right] =0.
 \end{eqnarray*}
 
 \subsection{Balances of mass and entropy}
 
  For the  mass density, we get by successive derivations
 \begin{equation*}
 	\left\{
 	\begin{array}{l}
 		\displaystyle\frac{\partial \rho }{\partial t}  + \rm{div}\left( \rho \,\boldsymbol{u}%
 		\right) =0, \\
 		\\
 		\displaystyle \frac{\partial (\nabla \rho)}{\partial t}  +
 		\rm{div}\left[ \nabla \left(
 		\rho \,\boldsymbol{u}\right) \right] =0, \\
 		\displaystyle \qquad\qquad\qquad\vdots \\
 		\displaystyle \frac{\partial (\nabla ^{n}\rho)}{\partial t}  +
 		\rm{div}\left[ \nabla ^{\it n}\left( \rho \,\boldsymbol{u}\right)
 		\right] =0,
 	\end{array}%
 	\right.
 \end{equation*}
 where we recall that,
 \begin{equation*}
 	\nabla \rho =\frac{\partial\rho}{\partial\boldsymbol{x}},\ \
 	\nabla (\rho\,\boldsymbol{u})
 	=\frac{\partial(\rho\,\boldsymbol{u})}{\partial\boldsymbol{x}},\ \ldots \ ,
 	\nabla^n \rho =\frac{\partial^n\rho}{\partial\boldsymbol{x}^{\,n}},\ \
 	\nabla^n (\rho\,\boldsymbol{u})
 	=\frac{\partial^n(\rho\,\boldsymbol{u})}{\partial\boldsymbol{x}^{\,n}}
 	.
 \end{equation*}
 If we assume $\left. \boldsymbol{\beta} _{1}\,\right\vert
 _{t=0}=\left. \nabla \rho\, \right\vert _{t=0}$, one can consider $\boldsymbol{\beta} _{1}=\nabla \rho $ as an independent
 variable. That is the same
 for $\boldsymbol{\beta} _{p}=\nabla ^{p}\rho $ with $\left.
 \boldsymbol{\beta} _{p}\,\right\vert _{t=0}=\left. \nabla ^{p}\rho\,
 \right\vert _{t=0}$. Then, all the previous equations are
 compatible with the mass
 conservation. But,%
 \begin{eqnarray*}
  \nabla \left( \rho \boldsymbol{u}\right)   & =&
 \nabla\rho\,  \otimes\, 	\boldsymbol{u} +\rho \  \nabla \,\boldsymbol{u}   \\
 	& \vdots & \\
 	 \nabla ^{n}\left( \rho \,\boldsymbol{u}\right)   &=&  \nabla
 	^{n}\rho  \,  \otimes\,
 	\boldsymbol{u}+C_{n}^{1}\ \nabla ^{n-1}\rho \,  \otimes\, \nabla \boldsymbol{u}+ 
 \ldots + {C}_{n}^{p}\,\nabla ^{n-p}\rho \,  \otimes\, \nabla ^{p}%
 	\boldsymbol{u}+ \ldots + \rho \, \nabla ^{n}\boldsymbol{u},
 \end{eqnarray*}
 Then,
 \begin{equation*}
 	\left\{
 	\begin{array}{l}
 		\displaystyle\frac{\partial \rho }{\partial t}+\rm{div}\left( \rho\, \boldsymbol{u}%
 		\right) =0, \\
 		\\
 		\displaystyle \frac{\partial \nabla \rho }{\partial
 			t}+\rm{div}\left[ \, \nabla\rho\,  \otimes\, 	\boldsymbol{u} +\rho \  \nabla \,\boldsymbol{u} \right] = 0, \\
 		\displaystyle \qquad\qquad\qquad\vdots \\
 		\displaystyle \frac{\partial \nabla ^{n}\rho }{\partial
 			t}+\rm{div}\left[\nabla
 		^{n}\rho  \,  \otimes\,
 	\boldsymbol{u}+C_{\it n}^{1}\ \nabla ^{{\it n}-1}\rho \,  \otimes\, \nabla \boldsymbol{u}+ 
 \ldots + {C}_{\it n}^{\it p}\,\nabla ^{\it n-p}\rho \,  \otimes\, \nabla ^{\it p}%
\boldsymbol{u}+ \ldots + \rho \, \nabla ^{\it n}\boldsymbol{u}\right] =0.
 	\end{array}%
 	\right.
 \end{equation*}
It is the same for the volumetric entropy   if we consider $
 \eta, \nabla \eta, \dots, \nabla^n \eta $ as independent
 variables.   
 If we note that
 \begin{equation*}
 	\frac{\partial \mathcal{P} }{\partial {\boldsymbol{\Phi
 		}}_{p}}=\nabla ^{p}\rho ,\quad \frac{\partial \mathcal{P} }{\partial
 		{\boldsymbol{\Psi }}_{p}}=\nabla ^{p}\eta , 
 \end{equation*}
we get,
 \begin{eqnarray*}
 	C_{n} &=&\left(\frac{\partial \mathcal{P} }{\partial {\boldsymbol{\Psi }}_{\it 1}}\,\vdots\, \frac{%
 		\partial {\boldsymbol{\Psi }}_{\it 2}}{\partial \boldsymbol{x}}\right)+\left(\frac{\partial
 		\mathcal{P} }{\partial {\boldsymbol{\Psi }}_{\it 2}}\,\vdots\,\frac{\partial {\boldsymbol{%
 				\Psi }}_{\it 3}}{\partial \boldsymbol{x}}\right)-\left(\frac{\partial \mathcal{P} }{\partial {%
 			\boldsymbol{\Psi }}_{\it 1}} \,\vdots\, \rm{div}\frac{\partial {\boldsymbol{\Psi }}%
 		_{\it 3}}{\partial \boldsymbol{x}}\right)+ \ldots   
 	+\left(\frac{\partial \mathcal{P} }{\partial {\boldsymbol{\Psi }}_{n-1}}\,\vdots\,\frac{%
 		\partial {\boldsymbol{\Psi }}_{n}}{\partial \boldsymbol{x}}\right)-\left(\frac{\partial
 		\mathcal{P} }{\partial {\boldsymbol{\Psi }}_{n-2}}\,\vdots\, \rm{div}\frac{\partial {%
 			\boldsymbol{\Psi }}_{\it n}}{\partial \boldsymbol{x}}\right)+ \ldots \\
 	&+&  (-1)^{p-1}\left(\frac{\partial \mathcal{P} }{\partial {\boldsymbol{\Psi
 		}}_{n-p}}\,\vdots\, {\rm{div}}_{p-1}\frac{\partial {\boldsymbol{\Psi
 		}}_{n}}{\partial
 		\boldsymbol{x}}\right)+{\ldots }+(-1)^{n-2}\left(\frac{\partial \mathcal{P} }{\partial {%
 			\boldsymbol{\Psi }}_{\it 1}}\,\vdots\,{\rm{div}}_{n-2}\frac{\partial {\boldsymbol{%
 				\Psi }}_{n}}{\partial \boldsymbol{x}}\right),
 	\\
 	D_{n} &=&\left(\frac{\partial \mathcal{P} }{\partial {\boldsymbol{\Phi }}_{\it 1}}\,\vdots\, \frac{%
 		\partial {\boldsymbol{\Phi }}_{\it 2}}{\partial \boldsymbol{x}}\right)+\left(\frac{\partial
 		\mathcal{P} }{\partial {\boldsymbol{\Phi }}_{\it 2}}\,\vdots\,\frac{\partial {\boldsymbol{%
 				\Phi }}_{\it 3}}{\partial \boldsymbol{x}}\right)-\left(\frac{\partial \mathcal{P} }{\partial {%
 			\boldsymbol{\Phi }}_{\it 1}} \,\vdots\, \rm{div}\frac{\partial {\boldsymbol{\Phi }}%
 		_{\it 3}}{\partial \boldsymbol{x}}\right)+ \ldots   
 	+\left(\frac{\partial \mathcal{P} }{\partial {\boldsymbol{\Phi }}_{n-1}}\,\vdots\,\frac{%
 		\partial {\boldsymbol{\Phi }}_{n}}{\partial \boldsymbol{x}}\right)-\left(\frac{\partial
 		\mathcal{P} }{\partial {\boldsymbol{\Phi }}_{n-2}}\,\vdots\, \rm{div}\frac{\partial {%
 			\boldsymbol{\Phi }}_{\it n}}{\partial \boldsymbol{x}}\right)+{\ldots } \\
 	&+&  (-1)^{p-1}\left(\frac{\partial \mathcal{P} }{\partial {\boldsymbol{\Phi
 		}}_{n-p}}\,\vdots\, {\rm{div}}_{p-1}\frac{\partial {\boldsymbol{\Phi
 		}}_{n}}{\partial
 		\boldsymbol{x}}\right)+{\ldots }+(-1)^{n-2}\left(\frac{\partial \mathcal{P} }{\partial {%
 			\boldsymbol{\Phi }}_{\it 1}}\,\vdots\,{\rm{div}}_{n-2}\frac{\partial {\boldsymbol{%
 				\Phi }}_{n}}{\partial \boldsymbol{x}}\right) 
 \end{eqnarray*}
 and we obtain,
\begin{theorem}
The system of  equations of processes for multi-gradient fluids can be written in the
 divergence form :
 \begin{equation}
 	\label{System 6}\left\{
 	\begin{array}{l}
 		\displaystyle\frac{\partial }{\partial t}\left( \frac{\partial \mathcal{P} }{%
 			\partial R}\right) +\rm{div}\left [\frac{\partial (\mathcal{P}\,{\boldsymbol{u}} )}{\partial\it R}\right ] =0 \\
 		\displaystyle\frac{\partial }{\partial t}\left( \frac{\partial \mathcal{P} }{%
 			\partial {\boldsymbol{\Phi }}_{\it 1}}\right) +\rm{div}\left[ \frac{\partial
 			\left( \mathcal{P} \,\boldsymbol{u}\right) }{\partial {\boldsymbol{\Phi }}_{\it 1}}+%
 		\frac{\partial \mathcal{P} }{\partial\it R}\frac{\partial \boldsymbol{u}}{\partial {\boldsymbol{x}}}%
 		\right] =0 \\
 		\displaystyle\qquad\qquad\qquad\vdots \\
 		\displaystyle\frac{\partial }{\partial t}\left( \frac{\partial \mathcal{P} }{%
 			\partial {\boldsymbol{\Phi }}_{\it n}}\right) +\rm{div}\left[ \frac{\partial
 			\left( \mathcal{P} \,\boldsymbol{u}\right) }{\partial {\boldsymbol{\Phi }}_{\it n}}+
  {C}_{\it n}^{1}\left(\frac{\partial \mathcal{P} \,}{\partial {\boldsymbol{\Phi }}%
 			_{{\it n}-1}}\,\otimes\, \frac{\partial \boldsymbol{u}}{\partial {\boldsymbol{x}}}\right) + \ldots + 
 {C}_{\it n}^{\it p}\left( \frac{\partial \mathcal{P} \,}{\partial {\boldsymbol{\Phi }}%
 			_{\it n-p}}\, \otimes\, \frac{\partial ^{\it p}\boldsymbol{u}}{\partial {\boldsymbol{x}}^{\it p}}\right)+{\ldots} + 
 		\frac{\partial \mathcal{P} }{\partial \it R}\frac{\partial ^{n}\boldsymbol{u}}{\partial {\boldsymbol{x%
 			}}^{n}}\right] =0 \\
 		\displaystyle\frac{\partial }{\partial t}\left( \frac{\partial \mathcal{P} }{%
 			\partial \tilde T}\right) +\rm{div}\left[ \frac{\partial \left( \mathcal{P} \,\boldsymbol{%
 				u}\right) }{\partial \it\tilde 
 			  T}\right] =0 \\
 		\displaystyle\frac{\partial }{\partial t}\left( \frac{\partial \mathcal{P} }{%
 			\partial {\boldsymbol{\Psi }}_{\it 1}}\right) +\rm{div}\left[ \frac{\partial
 			\left( \mathcal{P} \,\boldsymbol{u}\right) }{\partial {\boldsymbol{\Psi }}_{\it 1}}+%
 		\frac{\partial \mathcal{P} }{\partial\it \tilde T}\frac{\partial \boldsymbol{u}}{\partial {\boldsymbol{x}}}%
 		\right] =0   \\
 		\displaystyle\qquad\qquad\qquad\vdots \\
 		\displaystyle\frac{\partial }{\partial t}\left( \frac{\partial \mathcal{P} }{%
 			\partial {\boldsymbol{\Psi }}_{\it n}}\right) +\rm{div}\left[ \frac{\partial
 			\left( \mathcal{P} \,\boldsymbol{u}\right) }{\partial {\boldsymbol{\Psi }}_{\it n}}+ 
 		 {C}_{\it n}^{1}\left( \frac{\partial \mathcal{P}}{\partial {\boldsymbol{\Psi }}%
 			_{{\it n}-1}}\,\otimes\, \frac{\partial \boldsymbol{u}}{\partial {\boldsymbol{x}}}\right)+ \ldots + 
 	 {C}_{\it n}^{\it p}\left( \frac{\partial \mathcal{P}}{\partial {\boldsymbol{\Psi }}%
 			_{\it n-p}}\,\otimes\,\frac{\partial ^{\it p}\boldsymbol{u}}{\partial {\boldsymbol{x}}^{\it p}}\right)+  \ldots + 
 		\frac{\partial \mathcal{P} }{\partial \it\tilde T}\frac{\partial ^{\it n}\boldsymbol{u}}{\partial {\boldsymbol{x%
 			}}^{\it n}}\right] =0 \\
 		\displaystyle\frac{\partial }{\partial t}\left( \frac{\partial \mathcal{P} }{%
 			\partial \boldsymbol{u}}\right) + \rm{div}\left[\frac{\partial \left( \mathcal{P}
 			\,\boldsymbol{u}\right) }{\partial \boldsymbol{u}} -\frac{\partial \mathcal{P} }{\partial\it \tilde T}\frac{\partial }{%
 			\partial \boldsymbol{x}}\left( {\ \boldsymbol{\Psi }}_{\it 1}-{\rm{div}}_{\it 2}{%
 			\boldsymbol{\Psi }}_{\it 2}+ \ldots + (-1)^{{\it n}-1}{\rm{div}}_{{\it n}-1}{\boldsymbol{%
 				\Psi }}_{{\it n}-1}\right)\right.\\
 		\displaystyle\ \quad \qquad \qquad\quad\left. -\frac{\partial \mathcal{P} }{%
 			\partial \tilde \mu}\frac{\partial }{\partial \boldsymbol{x}}\left( {\ \boldsymbol{%
 				\Phi }}_{\it 1}-{\rm{div}}_{\it 2}{\boldsymbol{\Phi }}_{\it 2}+ \ldots +(-1)^{n-1}{%
 			\rm{div}}_{n-1}{\boldsymbol{\Phi }}_{n-1}\right)
 		-C_{n}-D_{n}\right]=0,
 	\end{array}%
 	\right.
 \end{equation}%
\end{theorem} 
 
 \subsection{Symmetric form and stability of constant states}
 
 System (\ref{System 6}) admits constant solutions $(\rho _{e}, \eta _{e} 
 , \boldsymbol{u}_{e}$,  $\nabla \rho _{e}=0$, $\ldots$ ,   $\nabla^{n} \rho
 _{e}=0$,  $\nabla \eta _{e}=0$, $ \ldots$,  $\nabla^{n}{\eta
 	_{e}=0})$. Since the governing
 equations are invariant under Galilean transformation, we can assume that $%
 \boldsymbol{u}_{e}= 0 $.
 Near equilibrium, we look for the solutions of the linearized
 system which are proportional in the direction $\boldsymbol{k}$ to $\displaystyle e^{i\left(
 	x-\lambda t\right) }$,$\ $where $x$ is the scalar coordinate in
 this spread direction, $\lambda $ is a constant and $i^{2}=-1$. We
 denote $u$ as a scalar corresponding to the  velocity
 in the direction $\boldsymbol{k}$ of $\boldsymbol{u}$
 ($\boldsymbol{u}=u\ \boldsymbol{k}$).  We denote
 \begin{equation*}
 	\boldsymbol{U}  = \boldsymbol{U}_{{\it 0}}\,e^{i\left( x-\lambda  
 		t\right) } ,
 \end{equation*}
 the general form of the perturbations with
 \begin{equation*}
 	\boldsymbol{U}^{\star }=\left[ \,R, {\boldsymbol{\Phi }}%
 	_{\it 1}, {\ldots , \boldsymbol{\Phi }}_{n},\tilde T, {\boldsymbol{\Psi }}_{\it 1}, {\ldots ,%
 		\boldsymbol{\Psi }}_{n}, \boldsymbol{u}\right] \quad
\rm{ and}\quad\
 \boldsymbol{U}_{{\it 0}}^{\star }  = \left[ \,\it R_0, {%
 		\boldsymbol{\Phi }}_{10}, {\ldots ,\boldsymbol{\Phi }}_{n0}, \tilde T_{0}, {%
 		\boldsymbol{\Psi }}_{10},  \ldots , \boldsymbol{\Psi }_{n0}, \boldsymbol{u}_{0}%
 	\right].
 \end{equation*}
 We obtain 
 \begin{equation*}
 	\frac{\partial }{\partial t}\left( \frac{\partial \mathcal{P} }{\partial \boldsymbol{%
 			U}}\right) _{e}=\frac{\partial }{\partial \boldsymbol{U}}\left( \frac{%
 		\partial \mathcal{P} }{\partial \boldsymbol{U}}\right) _{e}\frac{\partial
 		\boldsymbol{U}}{\partial t}=-i\lambda \,\frac{\partial }{\partial
 		\boldsymbol{U}}\left( \frac{\partial \mathcal{P} }{\partial
 		\boldsymbol{U}}\right) _{e}\,\boldsymbol{U}_{\it 0}\ e^{i\left(
 		x-\lambda t\right) },
 \end{equation*}%
 where subscript $e$ means the values at equilibrium and we denote
 \begin{equation*}
 	\boldsymbol{G}\equiv \frac{\partial }{\partial \boldsymbol{U}}\left( \frac{%
 		\partial \,\mathcal{P} \boldsymbol{u}}{\partial \boldsymbol{U}}\right) _{e}^{\star }.
 \end{equation*}%
From%
 \begin{equation*}
 	{\rm{div}}\left( \frac{\partial \,\mathcal{P} \boldsymbol{u}}{\partial \boldsymbol{%
 			U}}\right) =\frac{\partial }{\partial x}\left( \frac{\partial
 		\,\mathcal{P}
 		\boldsymbol{u}}{\partial \boldsymbol{U}}\right) ^{\star }=\frac{\partial }{%
 		\partial \boldsymbol{U}}\left( \frac{\partial \,\mathcal{P} \boldsymbol{u}}{\partial
 		\boldsymbol{U}}\right) ^{\star}\frac{\partial
 		\,\boldsymbol{U}}{\partial x},
 \end{equation*}%
 we get,%
 \begin{equation*}
 	{\rm{div}}\left( \frac{\partial \,\mathcal{P} \boldsymbol{u}}{\partial \boldsymbol{%
 			U}}\right) _{e}=i\ \boldsymbol{G\ U}_{{\it 0}}\,e^{i\left( x-\lambda
 		t\right) }.
 \end{equation*}%
At equilibrium, $ \nabla \rho _{e}=0, \ldots ,\nabla^{n} \rho
 _{e}=0, $ $\nabla \eta _{e}=0, \ldots ,\nabla^{n}{\eta _{e}=0} $, which
 implies
 \begin{equation*}
 	\left( \frac{\partial \,\mathcal{P} \boldsymbol{u}}{\partial {\boldsymbol{\Phi }}_{\it 1}%
 	}\right) _{e} =0, \,{\ldots }\,,\,\left( \frac{\partial \,\mathcal{P}}{%
 		\partial {\boldsymbol{\Phi }}_n}\right) _{e}=0,\quad\left( \frac{\partial \,\mathcal{P}}{\partial {\boldsymbol{\Psi }}_{\it 1}}\right) =0, \, \ldots\,,\,\left( \frac{\partial \,\mathcal{P} \boldsymbol{u}}{\partial
 		{\boldsymbol{\Psi }}_n}\right) _{e}=0,
 \end{equation*}%
 \begin{equation*}
 	\left( \frac{\partial \,\mathcal{P} }{\partial R}\right) _{e}\frac{\partial ^{p}%
 		\boldsymbol{u}}{\partial x^{p}}=\rho _{e}\frac{\partial ^{p}\boldsymbol{u}}{%
 		\partial x^{p}}=i^{p}\rho _{e}\boldsymbol{\ u}_{{\it 0}} \,e^{i\left( x-\lambda
 		t\right) }\quad \mathrm{and}\quad \left( \frac{\partial \,\mathcal{P} }{\partial\tilde T}%
 	\right) _{e}\frac{\partial ^{p}\boldsymbol{u}}{\partial x^{p}}=\eta _{e}%
 	\frac{\partial ^{p}\boldsymbol{u}}{\partial x^{p}}=i^{p}\eta _{e}\boldsymbol{%
 		\ u}_{{\it 0}}\,e^{i\left( x-\lambda t\right) }.
 \end{equation*}%
 Due to%
 \begin{eqnarray*}
 	\left( \frac{\partial \mathcal{P} \,}{\partial R}\right) _{e}
 	&=&\rho _{e},\quad
 	\frac{\partial }{\partial x}\left[ {(-1)}^{p}{\rm{div}}_{p-1}{\boldsymbol{%
 			\Phi }}_{p}\right] ={(-1)}^{p}i^{p}\ {\boldsymbol{\Phi }}_{p\it 0}\ e^{i\left(
 		x-\lambda t\right) },\quad \\
 	\left( \frac{\partial \mathcal{P} \,}{\partial \tilde T}\right) _{e}
 	&=&\eta _{e},\quad
 	\frac{\partial }{\partial x}\left[ {(-1)}^{p}{\rm{div}}_{p-1}{\boldsymbol{%
 			\Psi }}_{p}\right] ={(-1)}^{p}i^{p}\ {\boldsymbol{\Psi }}_{p\it 0}\
 	e^{i\left( x-\lambda t\right) },
 \end{eqnarray*}%
 the last equation in system (\ref{System 6}) becomes%
 \begin{eqnarray*}
 	\frac{\partial }{\partial t}\left( \frac{\partial \mathcal{P} }{\partial \boldsymbol{%
 			u}}\right) +\rm{div}\left[ \frac{\partial \left( \mathcal{P} \,{\boldsymbol{u}}%
 		\right) }{\partial \boldsymbol{u}}\right]  - \rho _{e}\left( {\it i}^{\it 2}{%
 		{\boldsymbol{\Phi }}_{\it 1}}- {\it i}^{\it 3}{\boldsymbol{\Phi }}_{\it 2}+ \ldots +(-1) %
 	^{{\it n}+1}{\it i}^{{\it n}+1}{\boldsymbol{\Phi }}_{\it n}\right)  
 	 - \eta _{e}\left( {\it i}^{2}{\boldsymbol{\Psi }}_{\it 1}-{\it i}^{3}{\boldsymbol{\Psi }}%
 	_{\it 2}+{\ldots +(-1)}^{{\it n}+1}{\it i}^{n+1}{\boldsymbol{\Psi }}_{\it n}\right) =0
 	.
 \end{eqnarray*}%
 We denote%
 \begin{equation*}
 	\boldsymbol{A}=\frac{\partial }{\partial \boldsymbol{U}}\left( \frac{%
 		\partial \,\mathcal{P} }{\partial \boldsymbol{U}}\right) _{e}^{\star}.
 \end{equation*}%
 which is a symmetric matrix. From the relations
 \begin{equation*}
 	\rm{div}\left({\it i}^{\it p}\rho _{e}\boldsymbol{u}\right) ={\it i}^{{\it p}+1}\rho _{e}%
 	\boldsymbol{u}_{_{0}}e^{{\it i}\left( x-\lambda t\right) }={\it i}^{{\it p}+1}\rho _{e}%
 	\boldsymbol{u}\ \quad \mathrm{and}\quad \rm{div}\left({\it i}^{\it p}\eta _{e}%
 	\boldsymbol{u}\right) ={\it i}^{{\it p}+1}\eta _{e}\boldsymbol{u}_{_{0}}e^{{\it i}\left(
 		x-\lambda t\right) }={\it i}^{{\it p}+1}\eta _{e}\boldsymbol{u},
 \end{equation*}%
 System (\ref{System 6}) writes
 \begin{equation*}
 	\left\{
 	\begin{array}{l}
 		\displaystyle\frac{\partial }{\partial t}\left( \frac{\partial \mathcal{P} }{%
\ 			\partial \tilde\mu}\right) +\rm{div}\left[ \frac{\partial \left( \mathcal{P} \,\boldsymbol{%
 				u}\right) }{\partial  \tilde\mu}\right] =0 \\
 		\displaystyle\frac{\partial }{\partial t}\left( \frac{\partial \mathcal{P} }{%
 			\partial {\boldsymbol{\Phi }}_{\it 1}}\right) +\rm{div}\left[ \frac{\partial
 			\left( \mathcal{P} \,\boldsymbol{u}\right) }{\partial {\boldsymbol{\Phi }}_{\it 1}}%
 		\right] +{\it i}^{\it 2}\rho _{e}\boldsymbol{u}=0 \\
 		\displaystyle\qquad \qquad \vdots \\
 		\displaystyle\frac{\partial }{\partial t}\left( \frac{\partial \mathcal{P} }{%
 			\partial {\boldsymbol{\Phi }}_{n}}\right) +\rm{div}\left[ \frac{\partial
 			\left( \mathcal{P} \,\boldsymbol{u}\right) }{\partial {\boldsymbol{\Phi }}_{\it n}}%
 		\right] +{\it i}^{{\it n}+1}\rho _{e}\boldsymbol{u}=0 \\
 		\displaystyle\frac{\partial }{\partial t}\left( \frac{\partial \mathcal{P} }{%
 			\partial \tilde T}\right) +\rm{div}\left[ \frac{\partial \left( \mathcal{P} \,\boldsymbol{%
 				u}\right) }{\partial \tilde T}\right] =0 \\
 		\displaystyle\frac{\partial }{\partial t}\left( \frac{\partial \mathcal{P} }{%
 			\partial {\boldsymbol{\Psi }}_{\it 1}}\right) +\rm{div}\left[ \frac{\partial
 			\left( \mathcal{P} \,\boldsymbol{u}\right) }{\partial {\boldsymbol{\Psi }}_{\it 1}}%
 		\right] +{\it i}^{\it 2}\eta _{e}\boldsymbol{u}=0 \\
 		\displaystyle\qquad \qquad \vdots \\ \label{System  5}
 		\displaystyle\frac{\partial }{\partial t}\left( \frac{\partial \mathcal{P} }{%
 			\partial {\boldsymbol{\Psi }}_{n}}\right) +\rm{div}\left[ \frac{\partial
 			\left( \mathcal{P} \,\boldsymbol{u}\right) }{\partial {\boldsymbol{\Psi }}_{\it n}}%
 		\right] +{\it i}^{{\it n}+1}\eta _{e}\boldsymbol{u}=0 \\
 		\displaystyle\frac{\partial }{\partial t}\left( \frac{\partial \mathcal{P} }{%
 			\partial \boldsymbol{u}}\right) +\rm{div}\left[ \frac{\partial \left( \mathcal{P}
 			\,\boldsymbol{u}\right) }{\partial \boldsymbol{u}}\right] -\rho _{e}\left(
 	{\it i}^{\it 2}{\boldsymbol{\Phi }}_{\it 1}-{\it i}^{\it 3}{\boldsymbol{\Phi }}_{\it 2}+{\ldots +(-1)}%
 		^{{\it n}+1}{\it i}^{{\it n}+1}{\boldsymbol{\Phi }}_{\it n}\right) 
 	 -\eta _{e}\left( {\it i}^{\it 2}%
 		{\boldsymbol{\Psi }}_{\it 1}-{\it i}^{\it 3}{\boldsymbol{\Psi }}_{\it 2}+{\ldots +(-1)}%
 		^{{\it n}+1}{\it i}^{{\it n}+1}{\boldsymbol{\Psi }}_{\it n}\right) =0 ,
 	\end{array}%
 	\right.  
 \end{equation*}
 which can be written in the form
 \begin{equation}
 	- i\ \lambda ~\boldsymbol{A\ U}+i\ \boldsymbol{G\
 		U}+i^{\it 2}\boldsymbol{C\ U=0} ,\label{vp}
 \end{equation}%
 where $\boldsymbol{C}$ is a matrix with $2(n+1)+1$ lines and $%
 2(n+1)+1$ columns which can be written as,
 \begin{equation*}
 	\boldsymbol{C} = \left[
 	\begin{array}{ccccccccc}
 		0, & 0, & 0, & 0, & 0, & 0, & 0, & 0, & 0\  \\
 		0, & 0, & 0, & 0, & 0, & 0, & 0, & 0, & i^{2}\rho _{e} \\
 		\vdots & \vdots & \vdots & \vdots & \vdots & \vdots & \vdots &
 		\vdots &
 		\vdots \\
 		0, & 0, & 0, & 0, & 0, & 0, & 0, & 0, & i^{n+1}\rho _{e} \\
 		0, & 0, & 0, & 0, & 0, & 0, & 0, & 0, & 0 \\
 		0, & 0, & 0, & 0, & 0, & 0, & 0, & 0, & i^{2}\eta _{e} \\
 		\vdots & \vdots & \vdots & \vdots & \vdots & \vdots & \vdots &
 		\vdots &
 		\vdots \\
 		0, & 0, & 0, & 0, & 0, & 0, & 0, & 0, & i^{n+1}\eta _{e} \\
 		0, & -(-i)^{2}\rho _{e}, & {\ldots ,} & -(-i)^{n+1}\rho _{e}, & 0,
 		& -(-i)^{2}\eta
 		_{e}, & {\ldots ,} & -(-i)^{n+1}\eta _{e}, & 0%
 	\end{array}%
 	\right],
 \end{equation*}
 and  Eq. \eqref{vp}  becomes :
 \begin{equation*}
 	i\, \left( \boldsymbol{G}+i\ \boldsymbol{C}-\lambda
 	~\boldsymbol{A}\right) \boldsymbol{U}_{\it 0}\, e^{i\left( x-\lambda
 		t\right)}=0.
 \end{equation*}
 Due to $\overline{i\,\boldsymbol{C}}^{\;\star }=i\,\boldsymbol{C}$, matrix $%
 i\,\boldsymbol{C}$ is Hermitian  operator; consequently, $\boldsymbol{K}=\boldsymbol{G}+i\,\boldsymbol{C%
 }$ is also an Hermitian operator, but $\boldsymbol{A}$ is symmetric. The $\lambda$-roots of%
 \begin{equation*}
 	\left( \boldsymbol{K}-\lambda ~\boldsymbol{A}\right) \boldsymbol{U}_{\it 0}=0,
 \end{equation*}
 are the solutions of characteristic equation,
 \begin{equation*}
 	\det \left( \boldsymbol{K}-\lambda ~\boldsymbol{A}\right) =0,
 \end{equation*}
where $\boldsymbol{U}_{\it 0}$ is the eigenvector associated with eigenvalue $%
 \lambda $. Near an equilibrium state, and when  Legendre transformation  $\mathcal{P} $ of energy $E$ is  locally convex, $ 
 \boldsymbol{A}$ is a positive definitive matrix and the
 eigenvalues $\lambda $ are real. Consequently, 
 \begin{corollary}
 When $E$ is locally convex, perturbations
 $\boldsymbol{U}_{\it 0}\,e^{i \left( x-\lambda t\right)} $ are stable
 and the $\boldsymbol{U}$-form is   dispersive.
 \end{corollary}

 \section{Conclusion}

 We have extended the cases of  capillary fluids \cite{Gavrilyuk2,Gavage} to the most  general case of
 multi-gradient     fluids in density and volumetric entropy. These fluids can be  represented by an hyperbolic-parabolic system of equations.  
 The  
divergence form of governing equations implies a system  of Hermitian-symmetric equations  constituting the most general dispersive model of conservative fluids. The perturbations are stable in the domains where   the total volumetric  internal energy is a convex function of the main field of new variables. The multi-gradient fluids have    common properties with   simple systems of classical conservative fluids \cite{Gouin,Serrin}. Multi-gradient fluids correspond to fluid media typified by first integrals   represented by Kelvin's theorems    \cite{Gouin3}.

 {\bigskip }

\parindent 0pt { {\textbf{Acknowledgments}: The author   thanks the National Group of Mathematical Physics GNFM-INdAM for its support as visiting professor at the Department of Mathematics of University of Bologna.

{\bigskip }

{

\end{document}